%% file: scifile.tex
\documentclass[12pt]{article}

\usepackage{lineno, blindtext}

\RequirePackage{hyperref}
\RequirePackage{graphicx,xcolor,subcaption,siunitx,booktabs}
\RequirePackage{colortbl}
\RequirePackage{booktabs}
\RequirePackage{algorithm}
\RequirePackage[noend]{algpseudocode}
\RequirePackage{changepage}
\RequirePackage[twoside,%
				letterpaper,includeheadfoot,%
				layoutsize={8.125in,10.875in},%
                layouthoffset=0.1875in,%
                layoutvoffset=0.0625in,%
                left=38.5pt,%
                right=43pt,%
                top=43pt,
                bottom=32pt,%
                headheight=0pt,
                headsep=10pt,%
                footskip=25pt,
                marginparwidth=38pt]{geometry}
\RequirePackage[labelfont={bf,sf},%
                labelsep=period,%
                figurename=Fig.]{caption}
\RequirePackage{etoolbox}
\AtBeginEnvironment{tabular}{
\sffamily\fontsize{7.5}{10}\selectfont
}

\usepackage[utf8]{inputenc}

\newenvironment{sciabstract}{%
\begin{quote} \bf}
{\end{quote}}

\newcounter{lastnote}
\newenvironment{scilastnote}{%
\setcounter{lastnote}{\value{enumiv}}%
\addtocounter{lastnote}{+1}%
\begin{list}%
{\arabic{lastnote}.}
{\setlength{\leftmargin}{.22in}}
{\setlength{\labelsep}{.5em}}}
{\end{list}}
\title{COVID-19 Policy analysis: Labor structure dictates lockdown mobility behavior} 
\author
{Samuel Heroy$^{1,2\ast}$, Isabella Loaiza$^{3,4}$, Alex Pentland$^{3}$, Neave O'Clery$^{1,2}$,\\
\\
\normalsize{$^{1}$The Bartlett Centre for Advanced Spatial Analysis, University College London,}\\
\normalsize{London W1 4TJ, UK}\\
\normalsize{$^{2}$Mathematical Institute, University of Oxford,}\\
\normalsize{Radcliffe Observatory Quarter, Woodstock Road, Oxford OX2 6GG, UK}\\
\normalsize{$^{3}$MIT Media Laboratory, Massachusetts Institute of Technology,}\\
\normalsize{E15–383, 20 Ames Street, Cambridge, MA 02139, USA}\\
\normalsize{$^{4}$ ICTEAM Institute - Pôle en ingénierie mathématique, UCLouvain,}\\
\normalsize{Louvain-la-Neuve 1348, Belgium}\\
\\
\normalsize{$^\ast$Corresponding author; E-mail:  s.heroy@ucl.ac.uk.}}

\date{}

\begin{document} 
\maketitle

\input{abstract.tex}

\section{Introduction}
\input{introduction}

\begin{figure}[hbtp]
\centering
\includegraphics[width=13cm]{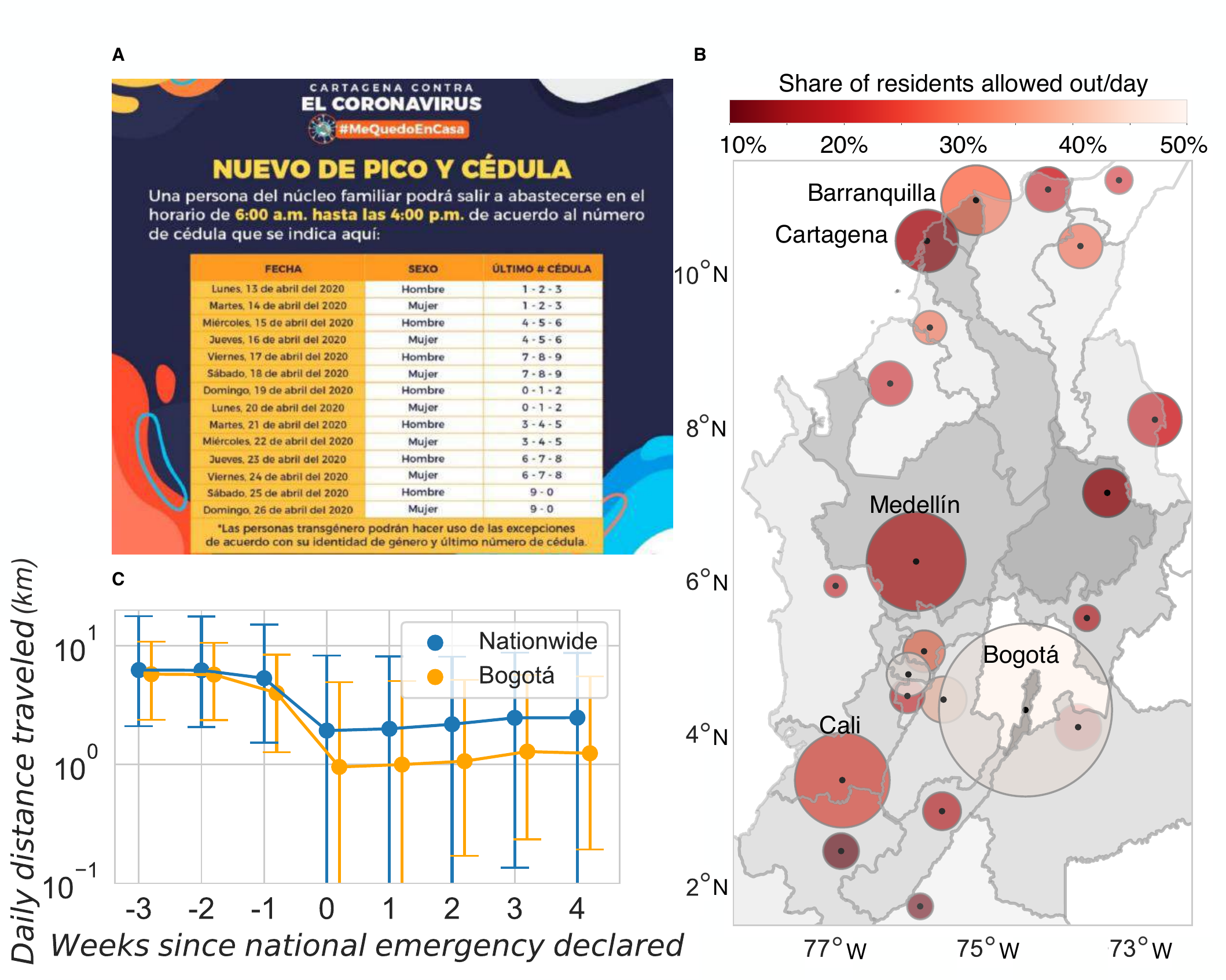}
\caption{\textbf{Colombian municipalities administered locally varying policies in which residents were allowed to go out for essential purposes on days corresponding to their national ID/gender.} A: During the study period, Medell\'in allowed out an average of 3 ID numbers daily ($30\%$ of residents). B: The capital municipalities (points, size proportional to population) of the most populous departments in Colombia, colored by the average share of residents allowed out per weekday from April 13-27. Departments are shaded by GDP per capita. C: Users nationwide reduced their mobility greatly from March 19 (end of week -1), when the government first announced national lockdown-related policies. From March 19-April 9 (weeks 0-2), residents nationwide were instructed to stay home unless absolutely necessary, but localized policies took effect from April 6 (week 2) in all municipalities except Bogot\'a, which implemented the pico y g\'enero policy on April 13 (week 3).}
\end{figure}
\section{Methods}
\input{methods}

\section{Results}
\input{Results/section1}

\input{Results/section2.tex}

\begin{figure}[hbtp]
\centering
\includegraphics[width=15cm]{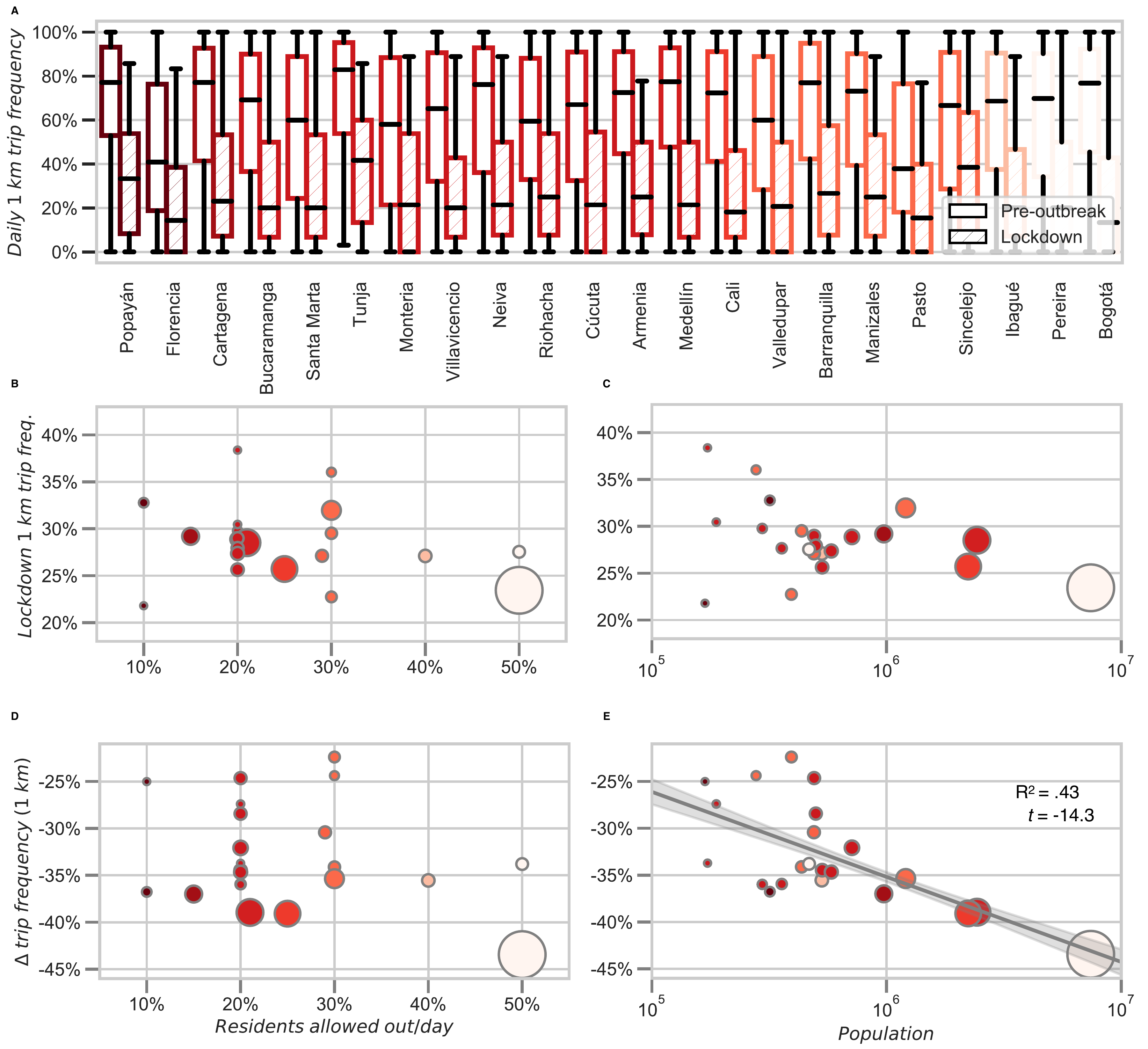}
    \caption{\textbf{City size, rather than mobility quota severity, is associated with more pronounced reduction in residents' trip frequency.} A: Empirical distributions of trip frequencies for detected residents in municipalities of interest (ordered by quota).
D-E: Municipality-level mean trip frequency (points, size proportional to population, color to mobility quota) during lockdown is not related significantly to quota or population. B-C: Increasing population (but not mobility quota) has a strong reducing effect on municipality-level median reduction in trip frequency (standard error bars and $95\%$ confidence intervals calculated via bootstrapping).}
\end{figure}

\begin{figure}[hbtp]
\centering
\includegraphics[width=15cm]{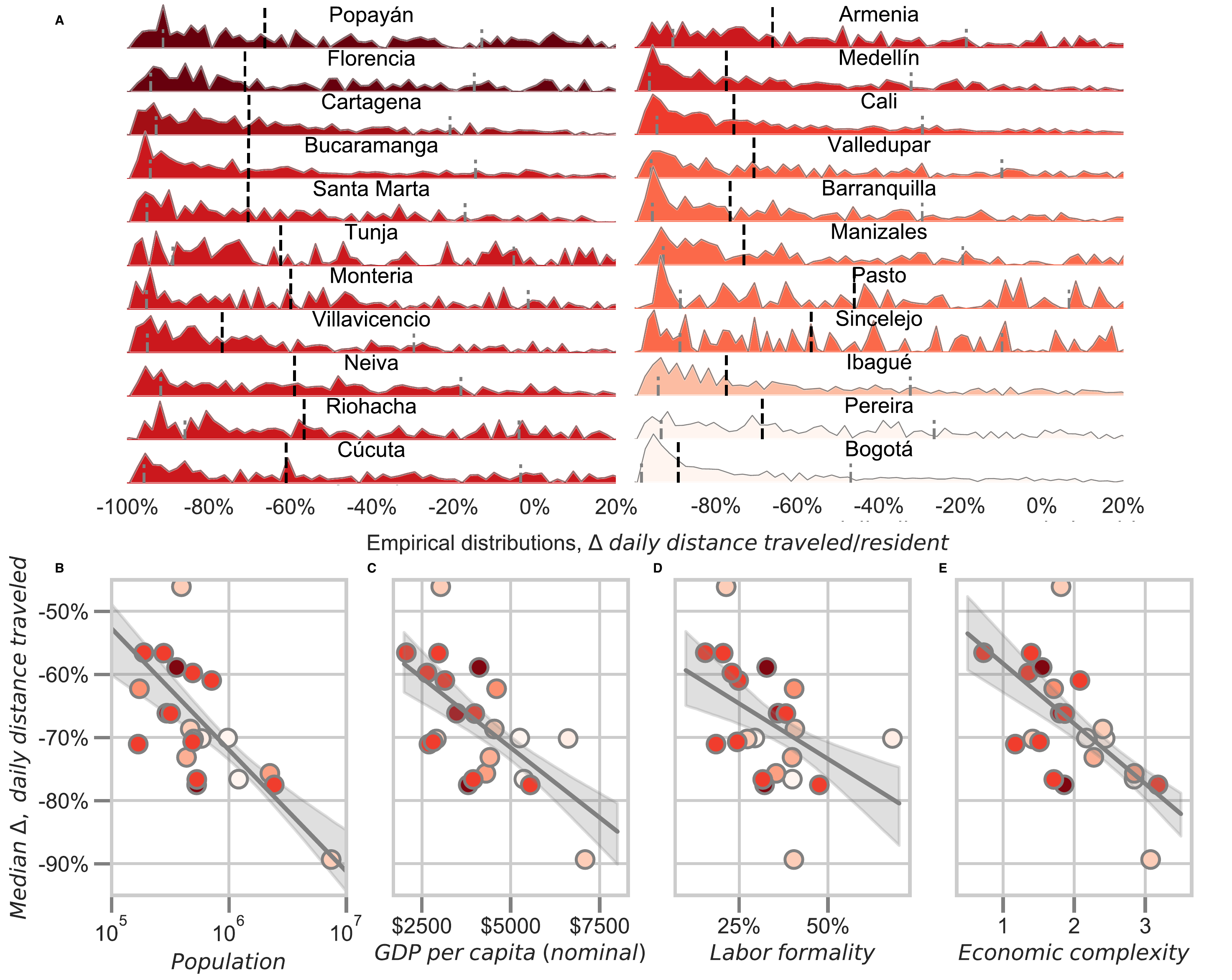}
\caption{\textbf{Residents of larger, wealthier municipalities with more formalized labor experienced more pronounced reductions in daily travel distance traveled during the lockdown period.} A. Empirical (kernel density) municipality-level distributions of relative change in daily distance traveled from the basal period to lockdown. Vertical dashed black (grey) lines represent municipality level median (quartile) estimates. Municipalities are colored/ordered by mobility quota (as in Fig. 2). B-E: Municipality population, as well as economic complexity, GDP per capita, and labor formality are associated with more pronounced relative change in daily distance traveled.}
\end{figure}

\input{Results/section3.tex}
\begin{figure}[hbtp]
\centering
\includegraphics[width=15cm]{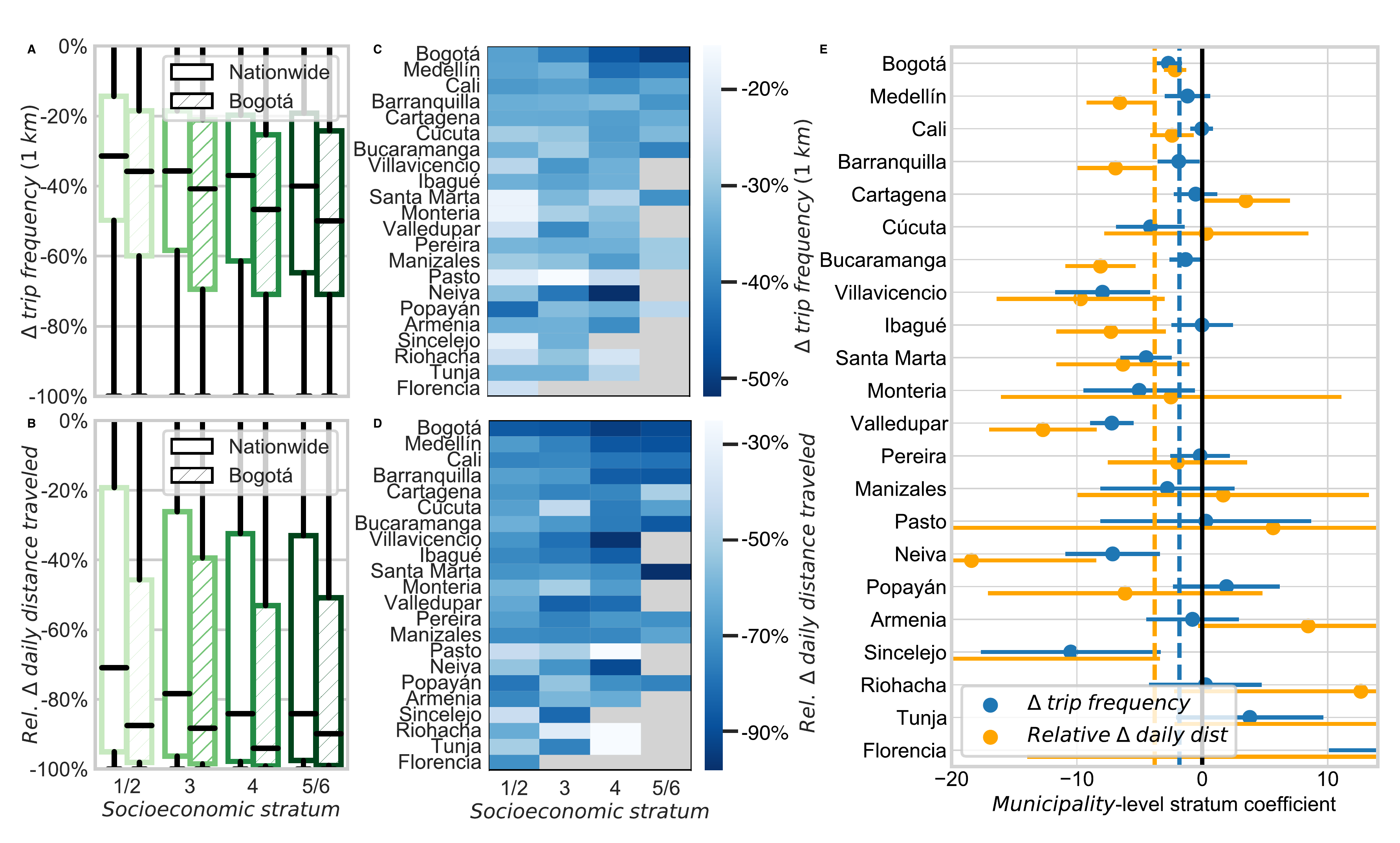}
\caption{\textbf{Within municipalities, higher socioeconomic status was generally associated with more pronounced mobility reduction.} A-B: Boxplot distributions of residential change in trip frequency and daily distance traveled for varying stratum. C-D: Municipality-level median estimates of change in trip frequency/daily distance traveled for varying stratum (municipalities ordered by population). E: Municipality-level coefficient estimates for the effect of stratum on changes in trip frequency and daily distance traveled (errorbars represent $95\%$ confidence intervals). Dashed lines represent average effects (under the fixed effects model) over all municipalities.}
\end{figure}

\input{Results/section4.tex}
\begin{figure}[hbtp]
\centering
\includegraphics[width=17cm]{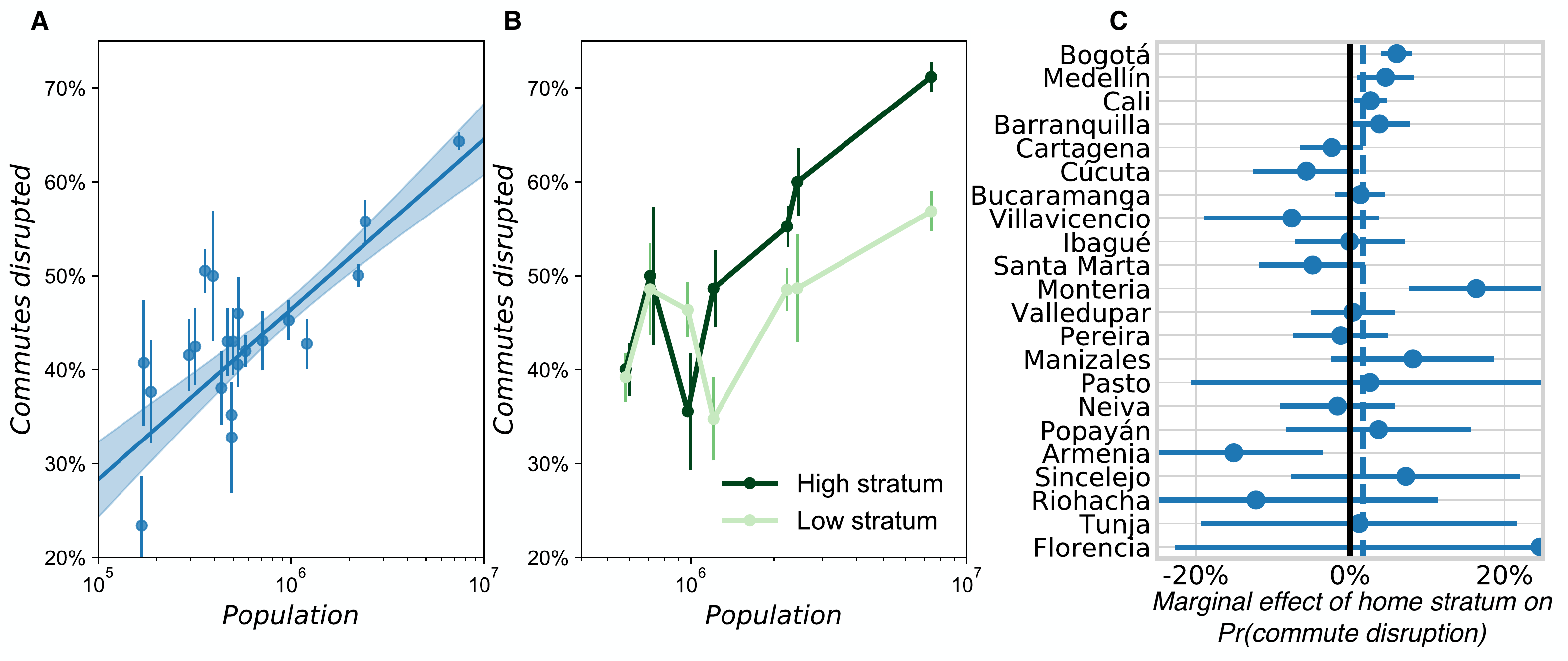}
\caption{\textbf{Effects of city size and socioeconomic stratum on commute disruption rate.} A: Effect of \emph{log(population)} on estimated \emph{commute disruption rate} (errorbars represent standard errors, $95\%$ confidence intervals for the OLS best fit line are calculated via bootstrapping). B: Point estimates of low/high \emph{stratum} subsample estimates of \emph{commute disruption} rate for the seven largest municipalities. C: Municipality-level average marginal effects of \emph{home stratum} on the likelihood a user has their commute disrupted, with point estimates/$95\%$ confidence intervals estimated via logistic regression. Dashed vertical lines represent average effects across all municipalities, which are ordered by population.}
\end{figure}

\begin{figure}[hbtp]
\centering
\includegraphics[width=15cm]{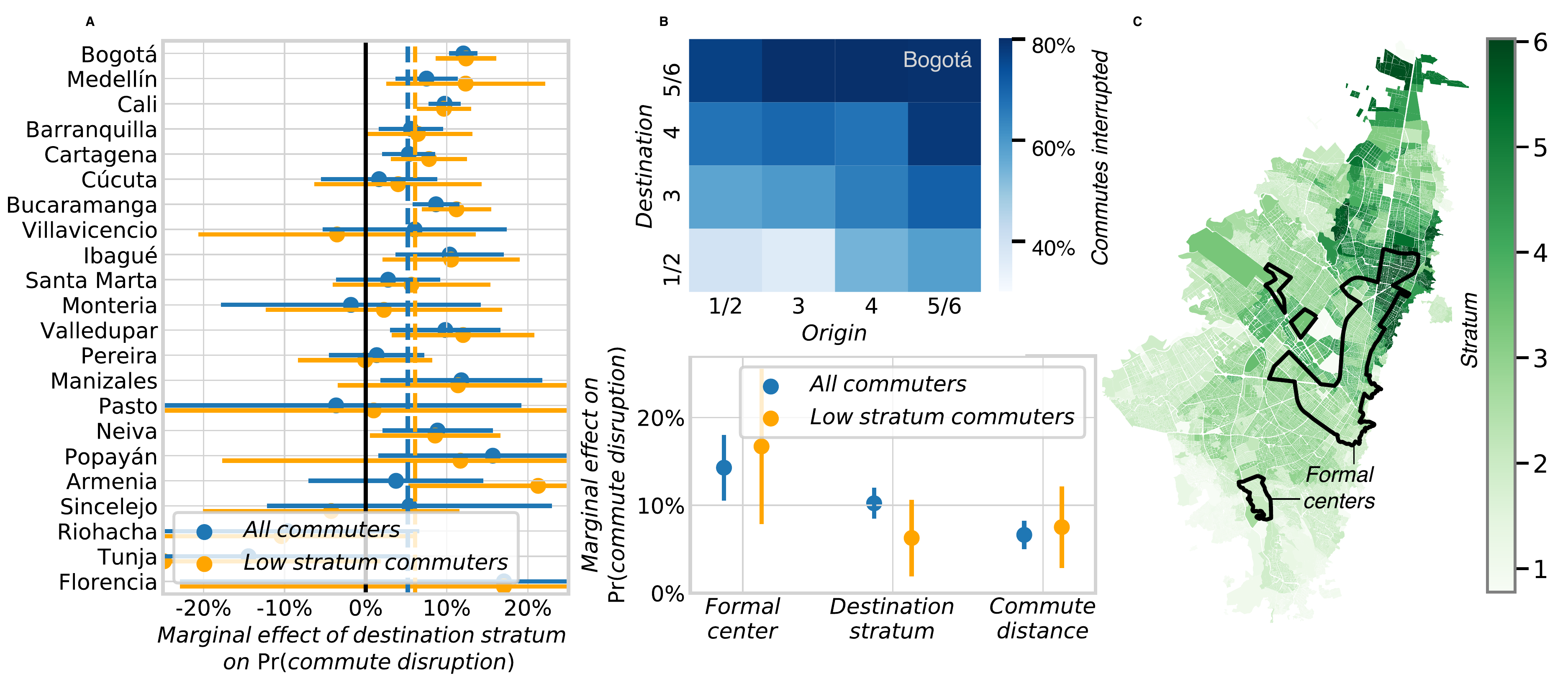}
\caption{\textbf{Commutes terminating in wealthier areas and formal labor centers were more likely to be disrupted.} A: Marginal effects of \emph{destination stratum} on the likelihood a user has their commute disrupted across all detected commuters/in low \emph{stratum} commuter subsamples of each municipality, ordered by population (point estimates/$95\%$ confidence intervals estimated via logistic regression). B: Estimates of commute disruption rate in Bogot\'a for varying home/work \emph{stratum}. C: Census-block characterizations of socioeconomic stratum in Bogot\'a, with the outlined regions representing formal employment centers - zones with at least 20,000 formal employees/km$^2$. D: Average marginal effects of commutes destination characteristics on the likelihood a commute is disrupted in Bogot\'a.}
\end{figure}
\section{Discussion}
\input{Discussion.tex}

\bibliography{scibib.bib}

\bibliographystyle{vancouver}

\begin{scilastnote}
\item 
The authors would like to thank in no particular order YY Ahn (Northwestern University), Mohsen Bahrami (MIT Media Lab), Morgan Frank (University of Pittsburgh), Eduardo Lora (Harvard University), Juan Camilo Chaparro (Universidad EAFIT), Julian Cristia (Inter-American Development Bank), Jaime Alfredo Bonet-Morón (Banco de la República Colombia), and Carlos Alberto Medina-Durango (Banco de la República Colombia) for helpful discussion. SH and NO acknowledge support from the PEAK Urban programme, funded by UKRI’s Global Challenge Research Fund, Grant Ref: ES/P011055/1.
\end{scilastnote}

\end{document}

%% file: abstract.tex
\begin{sciabstract}
Abstract: Countries and cities around the world have resorted to unprecedented mobility restrictions to combat Covid-19 transmission. Here we exploit a natural experiment whereby Colombian cities implemented varied lockdown policies based on ID number and gender to analyse the impact of these policies on urban mobility. Using mobile phone data, we find that the restrictiveness of cities’ mobility quotas (the share of residents allowed out daily according to policy advice) does not correlate with mobility reduction Instead, we find that larger, wealthier cities with more formalized and complex industrial structure experienced greater reductions in mobility. Within cities, wealthier residents are more likely to reduce mobility, and commuters are especially more likely to stay home when their work is located in wealthy or commercially/industrially formalized neighbourhoods. Hence, our results indicate that cities’ employment characteristics and work-from-home capabilities are the primary determinants of mobility reduction. This finding underscores the need for mitigations aimed at lower income/informal workers, and sheds light on critical dependencies between socioeconomic classes in Latin American cities.
\end{sciabstract}
Keywords: COVID-19, mobile device location data, human mobility, labour economics, inequality, policy analysis

%% file: introduction.tex
Across the world, governments and scientists alike have struggled immensely with the question of which policies will most effectively reduce the spread of COVID-19. In a cross-country analysis of case data, Hsiang \emph{et al.} find that in six countries various policies ranging from full lockdown to paid sick leave prevented or delayed an estimated 495 million cases from January-June 2020, but that the effects of specific policies differed from country to country \cite{hsiang2020effect}. Evidence from China \cite{fang2020human,maier2020effective} and across Europe \cite{flaxman2020estimating} demonstrates clear evidence that full and persistent lockdown is by far the most effective measure in curbing spread. However, it is not yet clear how local economic conditions affect policy success. 

Epidemic mitigation policies are not implemented in a vacuum. For instance, a tech hub has many jobs that can be performed remotely with relative ease, while a manufacturing center does not. Globally, cities with many teleworkable jobs have been better able to reduce work commutes \cite{sostero2020teleworkability, brussevich2020will}. More generally, wealthier cities and wealthier neighborhoods have been more adept at reducing urban mobility \cite{bonaccorsi2020economic, fraiberger2020uncovering, carvalho2020tracking, LUXURY}, and as well experienced lower Covid-19 death rates at least in the UK and USA \cite{ONS_data,zhang2020spatial}. However, the relationship between lockdown severity, city wealth, and observed mobility reduction remains not well understood. For example, are harsh mobility restrictions as effective in less wealthy cities? Or does the nature of the local economy outweigh policy severity?

Latin American countries generally had more time to prepare for the in-coming pandemic and develop appropriate policy measures \cite{lancet2020covid}. In Colombia, Ecuador, and Panamá, residents were allowed out for essential trips on days of the week corresponding to their national ID number and/or gender \cite{pc_spanish,pc_english}. We can quantify the restrictiveness of this type of policy via, i.e., the share of residents allowed out daily. We specifically focus on Colombian cities which implemented local variations of these policies. For example, in Bogot\'a residents were allowed out every other day based on their gender, whereas Florencia allowed just $10\%$ of residents out daily (Figure 1). Additionally, socioeconomic conditions vary significantly between Colombian cities, with the largest cities having much higher composite wealth and industry sophistication \cite{duranton2016agglomeration, lora2016path}. These variations in policy and wealth represent a natural experiment in which we examine the relationship between local policy, city wealth, and observed mobility reduction.   

We use mobile phone data (call records) to characterize changes in city residents’ urban mobility in an 11-day period beginning with the introduction of local lockdown measures. Urban mobility metrics from mobile phones have frequently been used to quantify mobility reductions in the wake of COVID-19 as well as other infectious diseases, both to characterize lockdown measures \cite{peak2018population, fang2020human}, and to predict epidemic spread \cite{badr2020association,xiong2020mobile, Gozzi2020.10.08.20204750, glaeser2020much}. 

Our key finding is that more severe mobility quotas have no significant impact on local mobility reduction levels. An implication of this finding is that fine-tuned restrictions, which calibrate the share of people allowed out daily, do not lead to a proportional decrease in mobility. On the other hand, city size is strongly correlated with mobility reduction, both in terms of trip frequency and daily distance traveled. Hence, cities with higher labor formality, GDP per capita, and industrial complexity experience more reduced mobility - irrespective of the quota imposed. Furthermore, within cities, commuters to wealthy and commercial areas are most disrupted. Hence, both high income and associated service workers are more likely to stay at home. Taken together, these results have important implications for the design of lockdown policies, the success of which depends critically on local economic conditions.

%% file: methods.tex
\subsection{Data}
\subsubsection{Municipal policy details}

 We collected (Spanish language) ``pico y cedúla"/``pico y género"  \cite{pc_english,pc_spanish} policy advice from official municipal websites and social media accounts for the municipalities of interest in order to compute mobility quotas. Generally, municipal advice specified days for which local residents with ID numbers ending in certain digits were allowed out (exceptions were made ofr key workers). In some cases, gender (or both gender/ID) was used to differentiate residents' allowed days out.
 
 In order to determine the share of residents allowed out daily (mobility quota), we make simple assumptions - that each municipality is equally represented by gender (noting that gender aspects of mobility have been studied using mobile phone data in Chile - \cite{gauvin2020gender} and in Panamá during the Covid-19 pandemic \cite{woskie2020men}) and that ID numbers are uniformly distributed (i.e. that residents in each municipality are equally likely to have an ID ending in any digit). With these assumptions, we first calculate the share of residents allowed out every day in late March/early April and then take the weekday average for the study period (April 13-27). We identify this average as the mobility quota $allowed_m$.
 
 In some cases, municipal advice included information regarding curfews (``toqué de queda'') or segmented days into times at which residents were allowed out. Also, some municipalities provided other information - including car-based restrictions (``pico y placá'') or information regarding where ``pico y cedúla'' would be implemented (e.g. at grocery stores and/or public transit). Finally, municipalities differed in weekend-based restrictions - in some municipalities, the local government administered a stay-at-home order on the weekend while in others there were either special restrictions for weekends or the same as those during the week. In accord with many other studies of both general mobility patterns \cite{louail2015uncovering,lotero2016rich} and social distancing \cite{carvalho2020tracking,fraiberger2020uncovering}, we focus our analysis on (more regular and predictable) weekday mobility patterns and therefore weekday restrictions as well.

\subsubsection{Mobility data}

CDR-extracted mobility measures correlate well with measures from other data sources \cite{alexander2015origin}. While XDR/GPS-based mobile phone data has more spatiotemporal accuracy, we use CDRs in part because of the wide penetration of our dataset throughout Colombia and in part because CDRs are reflective of users that contain both smartphones and ordinary cell phones. Although every mobility dataset may be subject to some bias \cite{wesolowski2013impact}, XDR/GPS-based data is likely to have more socioeconomic bias (without careful control) especially in a middle income country as such data typically comes exclusively from smartphones (in 2019, smartphone penetration in Colombia was only $59\%$ \cite{gsma}).

Our analysis relies on mapping individuals to spatial units - the granularity of these units is important to the the precision of our analysis. Generally, tower density varies according to municipal population, with some pairs of towers in densest cities being very close together. Pinpointing user location to very precise geographic locales (e.g. a single tower) with CDRs is generally difficult for a variety of both behavioral and technical (e.g. load-sharing) reasons, while accuracy may be enhanced by pinpointing to a more aggregate level \cite{pappalardo2020individual,flowminder}. We perform an agglomerative clustering algorithm \cite{scikit-learn} to join tower-associated Voronoi polygons with centroids that were within $100$ m of one another. Then, we use these clusters to perform a (more aggregate) Voronoi tessellation on tower cluster centers and associate each tower with its enclosing polygon.

\subsubsection{Home/work detection and active user selection}
\label{subsubsec:detection_selection}
Our analysis relies on having accurate estimations of home/work cell location, as well as spatio-temporal resolution of user location. Applying more stringent criteria for home/work cell detection and activity levels (activity refers to the frequency with which users make calls or receive in-network calls), however, reduces the sample size of our data and therefore the statistical power. Here, we describe the criteria that we use to transform raw CDRs to panel data. First, we expect to more accurately estimate true mobility levels for more active users - hence, we limit our subset to users who are active on a majority of weekdays during the lockdown period and at least 6 during a ``basal period" (January 1-March 15). 

There is considerable work on home cell detection from CDRs \cite{ahas2010using,louail2015uncovering,vanhoof2018assessing,pappalardo2020individual} - recent work suggests that identification is most robust when regularity of night-time location is used as a criterion and tower clustering is used  \cite{pappalardo2020individual}. Here, we identify a user's home with the cluster cell in which they are most consistently most active during night-time hours (here 0:00-7:00/22:30-24:00). Specifically, we identify a home cell if the user registers this cell as their most weeknight active night-time cell in at least three weeks during the basal period and a plurality of weeks during March 16-April 30 (inclusive of our lockdown study period). While it is commonplace to use night time location to define home cells (see for example \cite{louail2015uncovering,ahas2010using}), we focus on the consistency of their detected night-time location across weeks as our calculations depend on users keeping the same home cell during the basal/lockdown periods. This criterion is especially important here because it has been observed that many people have moved residence before/during lockdown \cite{escape} - residential movement during the study period would tend to inflate lockdown mobility. 

Additionally, the problem of work cell location is quite commonplace. We identify work cells in an analogous manner to our home detection approach \cite{louail2015uncovering}. We identify a user with a work cell if that cell is their most common day time (8:30-18:00) calling location on a plurality of the weeks in the basal period, and if they are active in the work cell on at least $20\%$ of their active days in this period.


With these identifications, we transform location data from raw CDRs to panel data, in which we have estimated the home location as well as mobility levels before/during lockdown for $15,347$ users across 22 municipalities. There are roughly 1 million unique users across 2020 in the raw CDR dataset, but many of these are not sufficiently active (especially during the lockdown period), fail to register a sufficiently consistent home cell, and/or are not based in one of the municipalities of interest\footnote{Only approximately 50,000 users register a common detected home cell for both the pre-lockdown and study period. Of these, approximately 17,000 have homes in the municipalities of interest - activity thresholds further limit the sample size of our panel set.} For $9,069$ of the $15,347$ users, we also estimate a pre-pandemic work cell that is unique from their home cell (in the other cases, either we do not detect a work cell or the work cell and home cell are the same).  
\subsubsection{Socioeconomic stratum}

The national government surveys urban residential dwelling conditions to assign socioeconomic strata \cite{DANE}. This data is provided in the form of survey indices mapped to geographic areas. Stratum tends to be highly spatially correlated due to socioeconomic segregation. We first compute average strata for the second most granular designation - ``secciónes urbanos" (urban sections), and then we use geographically weighted spatial averaging to assign strata to our tower cluster cells. Finally, we use $k=2$ nearest neighbor averaging \cite{2020SciPy-NMeth} to assign stratum to clusters that do not intersect urban sections (see Fig. S2). The distribution of stratum estimations across users is depicted in Fig. S1B.

\subsubsection{Municipality-level demographic/economic variables}

In addition to cell-level socioeconomic stratum, we use municipality-level official estimations of (a) population, (b) nominal GDP per capita, (c) labor formality, and (d) economic complexity. Estimates of (a) and (b) are available from DANE \cite{DANE} at \url{https://www.dane.gov.co/index.php/estadisticas-por-tema/demografia-y-poblacion} and \url{https://dane.maps.arcgis.com/apps/MapSeries/index.html?appid=71f231f4e31a40ec8796d559544e9103}. For (c), we use the same very simple methodology as in \cite{lora2016path}. That is, we simply associate labor formality with the (2017) count of formal employees in the municipality \cite{datlas} divided by the working age (15-64) population \cite{DANE}. This approach of course suffers from a number of limitations - (i) some of these employees will be based outside the municipality; (ii) a share of workers will be outside the working age population; and (iii) not everyone in the working population is in the workforce. Despite these limitations, it is a commonly used methodology to estimate labor formality and we expect that none of these assumptions will overwhelmingly bias our estimations for labor formality in any particular municipality. Finally, for (d) we use municipality-level industrial employment data \cite{datlas} and then compute economic complexity using the method of Hidalgo \emph{et al.} \cite{hidalgo2009building,hidalgo2007product,lora2016path}.
\subsection{Mathematical modeling}
In general, our mathematical modeling relies on individual-level regression analyses, though we additionally perform certain aggregate analyses as well (results are generally comparable). Where appropriate, we use a weighted approach so that all municipalities in our sample carry equal weight in our regressions. Because our response variables generally obey complex distributions, we generally use a regression bootstrapping scheme to generate confidence intervals, $p$-values, and standard errors. A detailed enumeration of our methodology can be found in SI Section 1. Generally, we use the subscript $i$ for individuals in our dataset and the subscript $m(i)$ to denote the city to which a resident belongs (for municipality-level variables) as well as $c(i)$ to denote the cell to which a resident belongs (for cell-level variables). We use the subscripts $m$ and $c$ to indicate municipality/cell-level variables.
\input{tables/variable_list.tex}

%% file: tables/variable_list.tex
\begin{table}[htbp]
\caption{ List of variable names}
\begin{center}
\begin{tabular}{|c|c|}
 \hline
 $trip frequency_i$ & $\%$ of weekdays in which a user makes a trip of $\ge1$ km\\\hline
 $daily distance traveled_i$ & Weekday-average maximum distance from home cell in which user $i$ is observed \\\hline
 $disrupted_i$ & Indicator of commute disrupted for user $i$ ($=1$ if their commute is disrupted)\\\hline
 $allowed_m$ & $\%$ of residents allowed out on average day in municipality $m$\\\hline
 $pop_m$ & Population of municipality $m$\\\hline
 $strat_c$ & Socioeconomic stratum of cell $c$\\\hline
$complex_m$ & Industrial complexity of municipality $m$\\\hline
$formal_m$ & Formality rate of municipality $m$\\\hline
$GDP_m$ & GDP per capita of municipality $m$\\\hline
$cell\;size_c$ & Area in square km of cell $c$\\\hline
$formal_c$ & Indicator of whether cell $c\in Bogot\acute{a}$ is formal\\\hline
$commute\;dist_i$ & Distance between home/work cells of commuter $i$\\\hline
$N$ & Number of individuals in sample\\\hline
$M$ & Number of municipalities in sample\\\hline
$N_m$ & Number of individuals in municipality $m$\\
\hline
\end{tabular}
\end{center}
\label{var_names}
\end{table}%

%% file: Results/section1.tex
\subsection{ID and gender-based mobility restrictions}

During the early stages of COVID-19 exposure, the Colombian government ordered municipal governments to impose local mobility restrictions. Most municipalities (22 departmental capitals) imposed ``pico y c\'edula” restrictions in which residents were allowed out on certain days corresponding to the terminating digit of their national ID numbers. However, the ID numbers allowed out daily varied between municipalities (Fig. 1A). For example, two of the smallest municipalities only allowed residents out roughly one weekday per every two weeks ($10\%$ of residents allowed out daily). The largest city and capital, Bogot\'a, allowed males/females out on alternating days ($50\%$ allowed out daily).

In order to quantify the restrictiveness of municipal policies, we collected local policy advice for the principle departmental capitals from government websites and/or social media. Whereas these ID/gender-based measures were advertised widely, and enforced in certain venues (e.g. public transit, access to banks/supermarkets), the extent of policy enforcement/adherence is generally unclear across municipalities, especially in the informal labor sector. 

For each departmental capital, we computed the average $\%$ of residents allowed out daily (or mobility quota) for 11 weekdays (April 13-27) in the pandemic's early stages (Fig. 1B). This corresponds to an initial 'lockdown' period in which all municipalities had recently enacted ID/gender-based restrictions. Interestingly, on average smaller municipalities permitted fewer people out daily (Pearson's $\rho=.43$, $p=.04$). Hence, with strong adherence, we would expect that residents of smaller municipalities traveled less frequently during lockdown. In the following section we will investigate whether this is observed in practice.

%% file: Results/section2.tex
\subsection{Effects of mobility quota and city size on lockdown mobility}

In order to evaluate the effects of localized mobility restrictions, we computed mobility indicators derived from call detail records (CDRs) for 22 out of 23 of Colombia's departmental capital municipalities (omitting one on account of low sample size). For each resident we compute \emph{trip frequency} which is the $\%$ of active days (days in which the user makes or receives a call) in which they travel 1 km from their home cell, and \emph{daily distance traveled} which is the average distance traveled from their home cell on active days. In order to assess changes in mobility relative to pre-lockdown, we also compute these metrics for a baseline period prior to lockdown and compute change measures. 

We find that most users across all municipalities reduced their daily distance traveled during lockdown (Fig. 1C). However, the extent of the reduction varied greatly across municipalities. We display municipality-level distributions of trip frequencies before and after lockdown in Figure 2A, and relative change (from baseline) in daily distance traveled in Figure 3A. Most municipalities had a mean lockdown trip frequency of $25-30\%$ (Fig. 2B-C). This represents a reduction in mobility from baseline levels that were $20-45\%$ higher, with the largest reduction in Bogot\'a (Fig 2D-E).

To what extent does mobility quota restrictiveness impact \emph{changes} in trip frequency and daily distance traveled relative to baseline levels? We find no evidence for an effect on daily distance traveled and statistically insignificant evidence ($p>.1$) for an effect on reduction in trip frequency. If a municipality reduced the amount of residents allowed out daily by $1\%$, residents on average decreased their trip frequency by only a meager additional $0.04\%$ (controlling for population, see Table~\ref{lockdown_freq_change}/\ref{daily_dist_trav_change} for detailed results).  

This finding - that more severe mobility quotas do not result in proportional reductions in mobility - is unexpected and important, especially considering the breadth of findings on the role of mobility restrictions in controlling infection spread \cite{kraemer2020effect, flaxman2020estimating, fang2020human}. We hypothesise that urban residents are more influenced by their economic capacity to comply with rules than by the precise measures implemented locally. As a first step, we investigate the role of urbanization (city size) in mobility reduction. An abundance of previous work has shown that city size is associated with economic prosperity - larger cities boast higher wages and GDP per capita, and a higher skilled as well as more formalized labor force working in more sophisticated sectors \cite{bettencourt2007growth,bettencourt2010urban, duranton2016agglomeration, lora2016path, davis2020comparative, brinkman2014supply,alves2015scale}. These factors generally improve residents' capacity to work remotely and reduce their mobility \cite{bonaccorsi2020economic, LUXURY, carvalho2020tracking, weill2020social,Gozzi2020.10.08.20204750}. 

We find that city size has a strong bearing on mobility reduction. Residents of larger municipalities had more pronounced reductions in trip frequency - for every tenfold increase in municipal population, the average resident reduced their trip frequency by an additional $9.07\%$. Additionally, a tenfold increase in population was associated with a $17.1\%$ additional reduction in daily distance traveled relative to the basal period. Moreover, when we aggregate to the municipality level, we find that city size explains nearly half of inter-municipality variance in both cases with $R^2=.43$ for change in trip frequency (Fig. 2E, Table ~\ref{lockdown_freq_change_mun}) and $R^2=.49$ for relative change in daily distance traveled (Fig. 3B, Table~\ref{lockdown_ddt_pool}). 

Hence, there is something about larger cities that better enables residents to reduce mobility despite less severe mobility quotas. This result may help to explain recent findings that initial COVID-19 growth is slower on a per-capita basis in larger cities \cite{ribeiro2020city}. We also find that aggregate levels of mobility reduction are significantly associated ($p<.01$) with municipality-level economic variables including GDP per capita and labor formality rate (Fig 3D-E, Table~\ref{econ_pool}), defined as the ratio of formal workers to the working age population \cite{lora2016path}. We construct a metric of industrial complexity \cite{hidalgo2009building, lora2016path,hidalgo2007product}, which is also, as expected, associated with reduction in daily distance traveled. We standardize these variables in order to facilitate comparison, and display coefficient estimates ordered from left to right in Table~\ref{econ_pool} (we also display results for unstandardized versions of these regressions as well as Spearman correlations in Supplementary Section 3.1). We find generally mixed results as to which of these (highly cross-correlated) variables best predict mobility reduction. While population is associated with the highest magnitude standardized coefficient estimate among all predictors for \emph{Relative} $\Delta$ \emph{daily dist. traveled}, GDP per capita has the highest in the case of $\Delta$ \emph{trip frequency}. Howver, for both mobility metrics, the confidence intervals for the predictor coefficients are mutually overlapping. More generally, it seems that economic advantages of larger cities favor mobility reduction and so we next investigate the effects of more granular (neighborhood-level) socioeconomic variables on mobility reduction.

\input{tables/regressions_sec2}
\input{tables/econvars}

%% file: tables/regressions_sec2.tex
\begin{table}[hbtp]
\scriptsize 
\caption{Individual/municipality-level regression estimates for effects of mobility quota and city size on lockdown behavior}
     \begin{subtable}{0.48\textwidth}
     \begin{center}
     \caption{\emph{Individual-level effects}}
     Dep. Variable: $\Delta\; trip\; frequency_i$\\
     \begin{tabular*}{\linewidth}{c  c  c  c  c}
        \toprule
        Ind. Variable & \textbf{(I)} & \textbf{(II)}  & \textbf{(III)}& \textbf{(IV)}\\ 
        
        \midrule
        $allowed_{m(i)}$  & -0.12*** &           &   0.04    &  0.04  \\
                           & (0.02)  &           &    (0.03) &   (0.03)\\
        $log(pop_{m(i)})$   &      &   -9.07***   &  -9.49***  &  -9.13***\\
                    &              &   (0.63)  &   (0.77)   &   (0.83)\\
        $Interaction_{m(i)}$ &              &           &            & -0.25\\
                   	&              &           &            &   (0.16)\\
        Const       & -29.97***   &   19.25***    &     20.74*** & 18.64*** \\
                    &  (0.67)     &     (3.79) &    (4.21) &  (4.65) \\ 
        
        \midrule
        N      (individuals)&15347   &  15347       &   15347           &  15347  \\
        $R^2$          &      0.002     &    0.018      & 0.018         & 0.018\\
        \bottomrule
     \end{tabular*}%
     \label{lockdown_freq_change}
     \end{center}
     \end{subtable}%
     \hspace*{\fill}%
     \begin{subtable}{0.48\textwidth}
     \begin{center}
     \caption{\emph{Aggregate effects}}
     Dep. Variable: $\mathrm{Mean}\left(\Delta\; trip\; frequency_{i\in m}\right)$\\
     \begin{tabular*}{\linewidth}{c  c  c  c  c}
         \toprule
         Ind. Variable & \textbf{(I)} & \textbf{(II)}  & \textbf{(III)}& \textbf{(IV)}\\ 
        
        \midrule
        $allowed_{m}$  &     -0.12***   &           &   0.04    &  0.04  \\
                      & (0.02)     &           &    (0.03) &   (0.03)\\
        $log(pop_{m})$   &      &   -9.07***   &  -9.49***  &  -9.13***\\
                    &              &   (0.64)  &   (0.79)   &   (0.86)\\
        $Interaction_m$ &              &           &            & -0.25\\
                   	&              &           &            &   (0.16)\\
        Const       & -29.97***   &   19.25***    &     20.74*** &18.64*** \\
                    &  (0.67)     &     (3.84) &    (4.31) &  (4.77) \\ 
        
        \midrule
         M      (municipalities)    &   22       &     22        &   22       &  22\\          
        $R^2$            &      0.05     &    0.43     & 0.43         & 0.44\\
        \bottomrule
     \end{tabular*}%
     \label{lockdown_freq_change_mun}
      \end{center}
     \end{subtable}
     \begin{subtable}{0.48\textwidth}
     \begin{center}
     \caption{\emph{Individual-level effects}}
     Dep. Variable: $Relative\;\Delta \; daily\; dist_i$\\
     \begin{tabular*}{\linewidth}{c  c  c  c  c}
        \toprule
        Ind. Variable & \textbf{(I)} & \textbf{(II)}  & \textbf{(III)}& \textbf{(IV)}\\ 
        
        \midrule
        $allowed_{m(i)}$  &     -0.3***   &           &   -0.1***    & -0.1**   \\
                      & (0.0)     &           &    (0.0) &   (0.0)\\
        $log(pop_{m(i)})$   &      &   -17.1***   &  -15.8***  &   -14.8***\\
                    &              &   (0.9)  &   (0.9)   &   (1.1)\\
        
        $Interaction_{m(i)}$ &              &           &            & -0.4**\\
                   	&              &           &            &   (0.2)\\
        Const       & -60.5***   &   29.4***    &     24.2*** &18.4*** \\
                    &  (1.0)     &     (5.3) &    (5.0) &  (6.2) \\ 
        
        \midrule
         N      (individuals)        &     15347        &     15347        &    15347         &  15347 \\   
        Pseudo $R^2$     &      0.001     &    0.002      & 0.002         & 0.003\\
        \bottomrule
     \end{tabular*}%
     \label{daily_dist_trav_change}
     \end{center}
     \end{subtable}%
     \hspace*{\fill}%
     \begin{subtable}{0.48\textwidth}
     \begin{center}
     \caption{\emph{Aggregate effects}}
     Dep. Variable: $\mathrm{Median}\left(Relative\;\Delta \; daily\; dist_{i\in m}\right)$\\
     \begin{tabular*}{\linewidth}{c  c  c  c  c}
        \toprule  
        Ind. Variable & \textbf{(I)} & \textbf{(II)}  & \textbf{(III)}& \textbf{(IV)}\\ 
        
        \midrule
        $allowed_{m}$  &     -0.3***   &           &   -0.1    &  -0.0  \\
                      & (0.1)     &           &    (0.1) &   (0.1)\\
        $log(pop_{m})$   &      &   -17.7***   &  -17.2***  &  -15.9***\\
                    &              &   (2.6)  &   (3.0)   &   (3.3)\\
        $Interaction_m$ &              &           &            & -0.9\\
                   	&              &           &            &   (0.6)\\
        Const       & -59.4***   &   34.5**    &     32.6* & 24.8 \\
                    &  (2.4)     &     (15.6) &    (16.8) &  (19.3) \\ 
        
        \midrule
         M      (municipalities)   &   22       &     22        &   22       &  22\\
        $R^2$                   &      0.12   &    0.49      & 0.49         & 0.51\\
        \bottomrule
     \end{tabular*}%
     \label{lockdown_ddt_pool}
     \end{center}
     \end{subtable}\\
     Statistical significance: *** indicates $p<.01$, ** indicates $.01\le p <.05$, * indicates $.05\le p<.1$. Parentheses indicate standard error estimates.
\end{table}

%% file: tables/econvars.tex
\begin{table}[hbtp]
\scriptsize 
\caption{Effects of standardized economic variables on lockdown behavior}
     \begin{subtable}{0.48\textwidth}
     \caption{}
     Dep. Variable: $\mathrm{Mean}\left(\Delta \; trip\; frequency_{i\in m}\right)$\\
     \begin{tabular*}{\linewidth}{c  c  c  c  c}
        \toprule
        Ind. variable & \textbf{(I)} & \textbf{(II)}  & \textbf{(III)}& \textbf{(IV)}\\ 

        \midrule
        $\widetilde{GDPpc_m}$   &    -4.13*** &           &       &   \\
                        &  (0.26)   &           &     &  \\
        $\widetilde{complex_m}$   &      &-3.94***      &    &  \\
                        &              &(0.26)     &      & \\
        $\widetilde{\log(pop_m)}$   &      &     &    -3.50*** & \\
                    &              &   &       (0.24) & \\
        $\widetilde{formality_m}$   &      &      &    & -3.46*** \\
                    &              &     &      &   (0.29)\\
        Const       & -32.89***   &   -32.89***    &   -32.89***   & -32.89***\\
                    &  (0.31)     &     (0.31) &   (0.31)   &  (0.31)  \\ 
        
        \midrule
        M      (municipalities)&22   &  22      &  22     & 22  \\
        $R^2$          &      0.60     &    0.54      & 0.43         & 0.42\\
        \bottomrule
     \end{tabular*}
     \label{econ_var_s1}
     \end{subtable}
     \hspace*{\fill}
     \begin{subtable}{0.48\textwidth}
     \caption{}
     Dep. Variable: $\mathrm{Median}\left(Relative\;\Delta \; daily\; dist_{i\in m}\right)$\\
     \begin{tabular*}{\linewidth}{c  c  c  c  c}
        \toprule
        Ind. variable & \textbf{(I)} & \textbf{(II)}  & \textbf{(III)}& \textbf{(IV)}\\ 

        \midrule
        $\widetilde{\log(pop_m)}$   &    -6.9*** &           &       &   \\
                        &  (1.0)   &           &     &  \\
         $\widetilde{complex_m}$   &           &    -5.9***       &       &   \\
                        &          &    (0.9)       &     &  \\
        $\widetilde{GDPpc_m}$   &            &    &   -5.7 &  \\
                                  &            &    &   (0.9)   & \\
        $\widetilde{formality_m}$   &      &      &    & -4.0*** \\
                    &              &     &      &   (1.1)\\
        Const       & -67.5***   &   -67.5***    &   -67.5***  & -67.5***\\
                    &  (1.2)     &     (1.2)  &    (1.2)   &  (1.2)   \\ 
        
        \midrule
        M      (municipalities)&22   &  22      &  22     & 22  \\
        $R^2$          &      0.49     &    0.37      & 0.33         & 0.17\\
        \bottomrule
     \end{tabular*}
     \label{econ_var_s2}
     \end{subtable}
     \label{econ_pool}
     $\widetilde{Variable}$ is used to indicate the enclosed variable is standardized across its distribution
\end{table}

%% file: Results/section3.tex
\subsection{Intra-city variation by socioeconomic status}

City size has a clear effect on mobility reduction, but - especially in Latin America - cities themselves are characterized by high internal income inequality \cite{roberts2005globalization}. Here we investigate how this drives mobility reduction at an intra-city level. Specifically, we ask if wealthier residents experienced higher levels of mobility reduction. 

We use the \emph{stratum} number assigned to each residential block as a proxy for residents' socioeconomic status \cite{DANE}. Stratum is a government-assigned designation corresponding to local housing conditions and quality, and has been frequently used to characterize socio-economic status \cite{medina2007stratification, lotero2016rich}. The stratum system encompasses 6 strata of progressively increasing socioeconomic status, with 1 signifying poor quality, often informal housing and 6 signifying the richest neighborhoods (stratum 6 is only present in select cities). Most residents countrywide live in medium stratum (2-3) housing.

In order to examine the effects of stratum on mobility reduction, we calculate changes in trip frequency and relative change in daily distance traveled as a function of stratum for each municipality (Fig. 4A-D). In general, wealthier residents experienced more pronounced reductions in both daily distance traveled and in trip frequency - in line with similar trends observed in cities across Latin America, Europe, the USA, and Asia \cite{Gozzi2020.10.08.20204750, fraiberger2020uncovering, weill2020social, LUXURY, carvalho2020tracking, bonaccorsi2020economic}. Pooling across municipalities (and taking into account fixed effects), we find that a one standard deviation increase in stratum was associated with a $1.82\%$ additional decrease in $\Delta$\emph{trip frequency (1 km)} and a $3.8\%$ additional decrease in \emph{Rel.} $\Delta$ \emph{daily distance traveled} ($p<.01$ in both cases, estimates are reported in Tables~\ref{tbl:stratum_effects_tf}/\ref{tbl:stratum_effects_ddt} and represented by dashed lines in Fig. 4E). The importance of stratum varies across municipalities - the effect of \emph{stratum} on $\Delta$ \emph{trip frequency} is significant ($p<.05$) in 10 out of 22 municipalities, and on \emph{Rel.} $\Delta$ \emph{daily distance traveled} is significant in 11 out of 22 municipalities (coefficient estimates displayed in Fig. 4E, tabulated in Supplementary Sec. 3.3). 

We have observed effects of socioeconomic inequality on mobility reduction at both an inter-city (previous section) and intra-city level. However, we now ask - are inter-city effects explained by differences in socioeconomic composition across cities? In other words, is there some effect of city size that is not simply captured by stratum variation? In order to disentangle these effects, we split each municipality into low ($1-2$) and high ($4-6$) stratum subsets and examine the effect of city size on mobility reduction across these subsets. For both groups, we find that increasing city size is associated with a greater reduction in trip frequency and daily distance traveled. The effect is comparably pronounced across low/high stratum groups - a tenfold increase in population is associated ($p<.01$) with an additional $7.14\%$(low)$/10.61\%$ (high) reduction in $\Delta$ \emph{trip frequency} (Tables~\ref{lockdown_freq_change_low}/\ref{lockdown_freq_change_hi})and an additional $14.3/23.5\%$ reduction in \emph{Rel.} $\Delta$ \emph{daily distance traveled} (Tables~\ref{daily_dist_trav_change_low}//\ref{daily_dist_trav_change_hi}). 

Overall, these findings underscore the role of city size in mobility reduction. We find that this effect is not limited to any socio-economic class in particular, but is instead consistent across strata. In other words, both the city in which a person resides and the socioeconomic status of their own neighborhood influence their level of mobility reduction. 
\input{tables/regressions_sec3a}

\input{tables/regressions_sec3}

%% file: tables/regressions_sec3a.tex
\begin{table}[hbtp]
\footnotesize 
\caption{Pooled/fixed effects of stratum on lockdown behavior}
     \begin{subtable}{0.48\textwidth}
     \caption{}
     Dep. Variable: $\mathrm{Mean}\left(\Delta\; trip\; frequency_{i\in m}\right)$\\
     \begin{tabular*}{\linewidth}{c  c  c  c  c}
        \toprule
        Ind. variable & \textbf{(I)} & \textbf{(II)}  & \textbf{(III)}& \textbf{(IV)}\\ 

        \midrule
        $strat_{c(i)}$  &    -4.00*** &           &   -2.38***    &   \\
                        &  (0.29)   &           &    (0.31) &  \\
        $\widetilde{strat}_{c(i)}$   &      &-1.82***      &    & -1.82*** \\
                    &              &(0.31)     &      &   (0.31)\\
        FE &          None    &None           &     Muni.       & Muni.\\
        Const       & -22.92***   &   -32.89***    &      & \\
                    &  (0.84)     &     (0.31) &     &   \\ 
        
        \midrule
        N      (individuals)&15347   &  15347       &   15347           &  15347  \\
        $R^2$          &      0.023     &    0.005      & 0.047         & 0.046\\
        \bottomrule
     \end{tabular*}
     \label{tbl:stratum_effects_tf}
     \end{subtable}
     \hspace*{\fill}
     \begin{subtable}{0.48\textwidth}
     \caption{}
     Dep. Variable: $Relative\;\Delta \; daily\; dist_i$\\
     \begin{tabular*}{\linewidth}{c  c  c  c  c}
        \toprule
        Ind. variable & \textbf{(I)} & \textbf{(II)}  & \textbf{(III)}& \textbf{(IV)}\\ 

        \midrule
        $strat_{c(i)}$  &    -6.1*** &           &   -4.5***    &   \\
                        &  (0.3)   &           &    (0.4) &  \\
        $\widetilde{strat}_{c(i)}$   &      &-4.2**      &    & -3.8*** \\
                    &              &(0.4)     &      &   (0.3)\\
        FE &          None    &None           &     Muni.       & Muni.\\
        Const       & -53.7***   &   -69.4***    &      & \\
                    &  (1.0)     &     (0.4) &     &   \\ 
        
        \midrule
        N      (individuals)&15347   &  15347       &   15347           &  15347  \\
        Pseudo $R^2$          &      0.002     &    0.001      & 0.005         & 0.005\\
        \bottomrule
     \end{tabular*}
     \label{tbl:stratum_effects_ddt}
     \end{subtable}
\end{table}

%% file: tables/regressions_sec3.tex
\begin{table}[hbtp]
\centering
\scriptsize 
\caption{Individual-level regression estimates for effects of mobility quota and city size on stratum-subset lockdown behavior}
     \begin{subtable}{0.48\textwidth}
      \begin{center}
     \caption{\emph{Low stratum subset}}
     Dep. Variable: $\Delta\; trip\; frequency_i$ \\
     \begin{tabular*}{\linewidth}{c  c  c  c  c}
        \toprule
        Ind. variable & \textbf{(I)} & \textbf{(II)}  & \textbf{(III)}& \textbf{(IV)}\\ 

        \midrule
        $allowed_{m(i)}$  & -0.04 &           &   0.09**    &  0.10**  \\
                           & (0.03)  &           &    (0.04) &   (0.04)\\
        $log(pop_{m(i)})$   &      &   -7.14***   &  -8.20***  &  -7.70***\\
                    &              &   (0.93)  &   (1.11)   &   (1.22)\\
       $Interaction_{m(i)}$ &              &           &            & -0.36\\
                   	&              &           &            &   (0.24)\\
        Const       & -30.85***   &   9.17    &     12.99** & 10.01 \\
                    &  (1.01)     &     (5.53) &    (6.02) &  (6.78) \\ 
        
        \midrule
        N      (individuals)&6593   & 6593       &  6593         &6593   \\
        $R^2$          &      0.000     &    0.011      & 0.012         & 0.013\\
        \bottomrule
     \end{tabular*}%
     \label{lockdown_freq_change_low}
     \end{center}
     \end{subtable}%
     \hspace*{\fill}%
     \begin{subtable}{0.48\textwidth}
      \begin{center}
     \caption{\emph{High stratum subset}}
     Dep. Variable:$\Delta\; trip\; frequency_i$  \\
     \begin{tabular*}{\linewidth}{c  c  c  c  c}
        \toprule
        Ind. variable & \textbf{(I)} & \textbf{(II)}  & \textbf{(III)}& \textbf{(IV)}\\ 

        \midrule
        $allowed_{m(i)}$  & -0.05 &           &   0.12**    &  0.13*  \\
                           & (0.06)  &           &    (0.06) &   (0.07)\\
        $log(pop_{m(i)})$   &      &   -10.61***   &  -11.96***  &  -11.70***\\
                    &              &   (2.15)  &   (2.24)   &   (2.90)\\
        $Interaction_{m(i)}$  &              &           &            & -0.16\\
                   	&              &           &            &   (0.52)\\
        Const       & -34.86***   &   25.20**    &     29.86** & 28.28* \\
                    &  (2.11)     &     (13.05) &    (13.20) &  (17.38) \\ 
        
        \midrule
        N      (individuals)&3551   &  3551      & 3551         &  3551    \\
        $R^2$    &      0.000     &    0.022      & 0.024       & 0.024\\
        \bottomrule
     \end{tabular*}%
     \label{lockdown_freq_change_hi}
     \end{center}
     \end{subtable}%
     
     \begin{subtable}{0.48\textwidth}
      \begin{center}
     \caption{\emph{Low stratum subset}}
     Dep. Variable: $Relative\;\Delta \; daily\; dist_i$ \\
     \begin{tabular*}{\linewidth}{c  c  c  c  c}
        \toprule
        Ind. variable & \textbf{(I)} & \textbf{(II)}  & \textbf{(III)}& \textbf{(IV)}\\ 
        \midrule
        $allowed_{m(i)}$  &     -0.2***   &         &   0.0    & 0.1**   \\
              & (0.0)     &           &    (0.0) &   (0.0)\\
        $log(pop_{m(i)})$   &      &   -14.3***   &  -14.4***  &   -10.0***\\
                &              &   (1.1)  &   (1.2)   &   (1.4)\\
    
       $Interaction_{m(i)}$  &              &           &            & -1.8***\\
               	&              &           &            &   (0.3)\\
        Const       & -63.1***   &   16.1**    &    16.2** &-10.4 \\
                &  (1.1)     &     (6.3) &    (6.7) &  (8.3) \\ 
    
    \midrule
     N      (individuals)   &   6593 &  6593   & 6593      & 6593  \\          
    Pseudo $R^2$     & 0.000  &   0.002      & 0.002         & 0.002\\
    \bottomrule
     \end{tabular*}%
     \label{daily_dist_trav_change_low}
     \end{center}
     \end{subtable}%
     \hspace*{\fill}%
     \begin{subtable}{0.48\textwidth}
     \begin{center}
     \caption{\emph{High stratum subset}}
     Dep. Variable: $Relative\;\Delta \; daily\; dist_i$ \\
     \begin{tabular*}{\linewidth}{c  c  c  c  c}
        \toprule
        Ind. variable & \textbf{(I)} & \textbf{(II)}  & \textbf{(III)}& \textbf{(IV)}\\ 

        \midrule
        $allowed_{m(i)}$  &     -0.3***   &       &   0.3***    & 0.2***   \\
              & (0.0)     &           &    (0.0) &   (0.0)\\
        $log(pop_{m(i)})$   &      &   -23.5***   &  -27.8***  &   -31.2***\\
                    &              &   (0.5)  &   (0.9)   &   (0.8)\\
        
        $Interaction_{m(i)}$  &              &           &            & 1.1***\\
                   	&              &           &            &   (0.2)\\
        Const       & -65.6***   &   65.2***    &    81.7*** &104.2*** \\
                    &  (1.0)     &     (3.4) &    (4.6) &  (4.8) \\ 
        \midrule
         N      (individuals)        &    3551   &  3551   &  3551  & 3551  \\
        Pseudo $R^2$   &  0.001     &    0.009      & 0.009         & 0.010\\
        \bottomrule
     \end{tabular*}%
     \label{daily_dist_trav_change_hi}
      \end{center}
     \end{subtable}%
     
\end{table}

%% file: Results/section4.tex
\subsection{Disruptions to work commutes}


Building on previous findings that show wealthier cities have more jobs that can be performed remotely \cite{dingel2020many, brussevich2020will, hatayama2020jobs}, we investigate how city size and socioeconomic stratum are linked to work commute disruptions. We examine whether (i) city size is linked to more commute disruptions; and (ii) whether residents' home/work stratum is linked to their tendency to forego their commutes.

As is standard practice, we identify home-work commutes via persistent origin-destination flows \cite{louail2015uncovering}. We define the commute disruption rate as the \% of flows that cease during lockdown (see SI Section 2.1 for results corresponding to other definitions). As with trip frequency and distance traveled, we find, as expected, that more commute flows are disrupted in larger cities (Fig. 5B, Table~\ref{table:logr_muni_disrupt}). For every tenfold increase in population, the municipality-level commute disruption rate rises $17.8\%$ ($p<.01$), while policy severity has no significant effect. 

Using a similar approach to that of the previous section, we examine whether higher stratum residents have more disrupted commute patterns (Fig. 5B-C, Table~\ref{table:logr_ind_disrupt}). In the largest 4 municipalities, we find at least marginal evidence ($p<.07$) that increasing stratum is associated with higher commute disruption probability. However, we only find significance at this level in only one other municipality, suggesting home stratum may play a less significant role in commute disruptions than on overall mobility reduction, particularly in smaller municipalities (the pooled average marginal effect under the fixed effects model is $1.6\%$, $p<.01$). In these cases, other factors such as the type or location of work may also drive commute disruption. 

While our data does not reveal the industry or occupation of commuters, we can distinguish workers according to where they work. Broadly, commuters to wealthier areas fall into two groups: they might either work at more formalized, sophisticated firms or provide services to wealthier employers or residents. In the latter case, evidence from the US points to faster and more pronounced drops in spending among wealthier classes during the pandemic \cite{chetty2020did}, corresponding to reduced demand for services in these areas. Here, we examine whether the stratum of a commuter's destination is linked to the likelihood that they disrupt their commute. Specifically, we ask whether low stratum commuters experienced more commute disruptions when working in high stratum or commercial/industrial areas.

As expected, we find that the likelihood that a commute is disrupted depends considerably on the destination stratum (Fig. 6). We find a significant ($p<.05$) effect of destination stratum on the probability a commute is disrupted in 11 out of 22 municipalities, including 6 of the largest 7 municipalities. These results hold across the general population as well as for the subset of low stratum ($\le2$) commuters (coefficient estimates are displayed in Fig. 6A and reported in Table~\ref{table:logr_ind_disrupt}/\ref{table:low_logr_ind_disrupt}) - pooled average marginal effects (under the fixed effects model) are quite similar for both cases ($5.2\%$ for all commuters, $6.1\%$ for low stratum residents, $p<.01$ in both cases). We see this effect most clearly in Bogot\'a, where commutes terminating in stratum 5/6 are twice as likely to be disrupted as commutes to stratum 5/6 (Fig. 6B). These results demonstrate that both low and high income workers that commuted to high income areas were much likelier to experience disrupted commutes.

\input{tables/disruptionslog}
\input{tables/logistic}
\input{tables/bogota}


Digging deeper into the characteristics of the commute destination, we use administrative formal employment data to identify commercial/industrial centers in Bogot\'a (Fig. 6C). These correspond to zones with at least $20,000$ formal jobs/km$^2$ \cite{guzman2017city}. Generally, these centers include a range of socio-economic strata, and include the central business district (Chico-Lago) and the government (La Candelaria). Controlling for commute distance, we find that commutes to these centers were on average $14.3\%$ ($p<.01$) more likely to be disrupted while a 1-stratum increase was associated with a $10.2\%$ ($p<.01$) increase in the probability a commute was disrupted (point estimates in Fig. 6D, results are reported in Table~\ref{table:bogota}/\ref{table:bogota_low})). As above, we find comparable effects when limiting our commuter sample to low stratum residents.

Hence, we find that higher income workers - with more tele-workable occupations - were more likely to discontinue their commutes. However, we also observe a sharp lockdown effect for work in higher income and more commercial locations. While we would certainly expect high income commuters to stay home, the evidence points towards a tendency for lower income workers who work in high commercial locations (presumably in service-oriented occupations) to also stay home. Therefore, we observe a lockdown ``trickle down'' effect - cities with higher income and more formalized firms not only have more commute disruptions for high income workers, but also for low income workers. This effect culminates in larger cities having higher levels of both commute and mobility disruption, trends that are consistent throughout our findings.

%% file: tables/disruptionslog.tex
\begin{table}[htbp]\centering
\caption{Effects of mobility quota and city size on commute disruption likelihood}
Dep. Variable: $\log\left(\frac{\mathrm{Pr}(disrupt_i=1)}{1-\mathrm{Pr}(disrupt_i=1))}\right)$\\
\begin{tabular}{c c c c c}
\toprule
Ind. variable & \textbf{(I)} & \textbf{(II)}  & \textbf{(III)}& \textbf{(IV)}\\ 

\midrule
$allowed_{m(i)}$  &     0.010***   &           &   -0.002 &  -0.000  \\
              & (0.002)     &           &    (0.02) &   (0.003)\\
$log(pop_{m(i)})$   &      &   0.745***    &  0.767***  &   0.767***\\
            &              &   (0.056)  &   (0.062)   &   (0.062)\\
$Interaction_{m(i)}$               &           &            & 0.031\\
           	&              &           &            &   (0.027)\\
Const       & -0.586***   &   -4.617***    &     -4.694*** & -4.764*** \\
            &  (0.055)     &     (0.323) &    (0.336) &  (0.342) \\ 
\midrule
Average marginal effects\\
\midrule
$allowed_{m(i)}$  &     0.2  &           &   -0.0 &  -0.0  \\
$log(pop_{m(i)})$   &      &   17.8    &  18.3  &   18.3\\
$Interaction_{m(i)}$ &      &       &   &   0.7\\
\midrule
 N       &     9069        &      9069         &     9069          &   9069  \\          
LL        &      -6153     &    -6075      & -6074        & -6074\\

\bottomrule
\addlinespace[1ex]
\end{tabular}
\label{table:logr_muni_disrupt}
\end{table}

%% file: tables/logistic.tex
\begin{table}[htbp]\centering
\caption{Logistic regression estimates for effect of home/work stratum on commute disruptions}
    \begin{subtable}{\textwidth}
    \begin{center}
    \caption{\emph{All commuters}}
    Dep. Variable: $\log\left(\frac{\mathrm{Pr}(disrupt_i=1)}{1-\mathrm{Pr}(disrupt_i=1))}\right)$\\
    \begin{tabular}{c c c c c c c c c}
    \toprule
    Ind. variable & \textbf{(I)} & \textbf{(II)}  & \textbf{(III)}& \textbf{(IV)} &\textbf{(V)} & \textbf{(VI)}  & \textbf{(VII)}& \textbf{(VIII)} \\ 
    
    \midrule
    $strat_{c(i,home)}$ &0.187***&&0.079***& 0.092***&    0.068***   &           &   0.031 &  0.046*  \\
                 &(0.021)&&(0.023)& (0.023)& (0.025)     &           &    (0.025) &   (0.026)\\
    $strat_{c(i,work)}$& &0.311***&0.282***& 0.289***  &      &   0.227***    &  0.222***  &   0.232***\\
                &&(0.021)&(0.023)&  (0.023)   &         &   (0.025)  &   (0.025)   &   (0.026)\\
    $strat_{c(i,home)}\times strat_{c(i,work)}$ &&&&-0.060***  &   &           &            & -0.043**\\
                      &&&&                       (0.018)&         &           &            &   (0.022)\\
    const    &    -0.817***  & -1.209***   &  -1.332***     &-1.365***&&&& \\
          & (0.059)    &     (0.064) &   (0.074)   &  (.075)&&&&   \\
    FE      &None&None&None& None & Muni.& Muni.&Muni. & Muni.\\
    \midrule
    Average marginal effects\\
    \midrule
    $strat_{c(i,home)}$ &4.5& &1.9 & 2.2 & 1.6 & & 0.7 & 1.1 \\
    $strat_{c(i,work)}$& &7.4 &6.7 &6.8 & &5.2 & 5.1 & 5.3\\
    $strat_{c(i,home)}\times strat_{c(i,work)}$ &&&&-1.4  &   &           &            & -1.0\\
    \midrule
     N    &9069&9069&9069&9069    &     9069        &      9069         &     9069          &   9069  \\          
    LL     &-6126&-6054&-6048&  -6042   &      -5953     &    -5915      & -5914        & -5912\\
    \bottomrule
    \addlinespace[1ex]
    \end{tabular}
    \label{table:logr_ind_disrupt}
    \end{center}
    \end{subtable}

    \begin{subtable}{\textwidth}
    \caption{\emph{Low stratum commuters}}
    \begin{center}
    Dep. Variable: $\log\left(\frac{\mathrm{Pr}(disrupt_i=1)}{1-\mathrm{Pr}(disrupt_i=1))}\right)$\\
    \begin{tabular}{c c c}
    \toprule
    Ind. variable & \textbf{(I)} & \textbf{(II)}   \\ 
    
    \midrule
    
    $strat_{c(i,work)}$& 0.326***&0.271***\\
                &(0.036)&(0.042) \\
    
    FE      &None& Muni.\\
    const    &    -1.255**  & \\
          & (0.106)    &      \\
    \midrule
    Average marginal effects\\
    \midrule
    $strat_{c(i,work)}$& 7.7 &6.1\\
    \midrule
     N    &3360 & 3360 \\          
    LL     &-2232 & -2153\\
    
    \bottomrule
    \addlinespace[1ex]
    \end{tabular}
    \label{table:low_logr_ind_disrupt}
    \end{center}
    \end{subtable}
    
\end{table}

%% file: tables/bogota.tex
\begin{table}[htbp]\centering
\caption{Effects of work stratum along with geographic labor formalization on commute disruption likelihood in Bogot\'a}
\begin{subtable}{\textwidth}
    \begin{center}
    \caption{\emph{All commuters}}
    Dep. Variable: $\log\left(\frac{\mathrm{Pr}(disrupt_i=1|i\in Bogot\acute{a})}{1-\mathrm{Pr}(disrupt_i=1| i \in Bogot\acute{a})}\right)$\\
    \begin{tabular}{c c c c c}
    \toprule
    Ind. variable & \textbf{(I)} & \textbf{(II)}  & \textbf{(III)}& \textbf{(IV)}\\ 
    
    \midrule
    $strat_{c(i,work)}$  &     0.562***   &           &     &  0.505***  \\
                  & (0.047)               &           &     &   (0.048)\\
    $formal_{c(i,work)}$      &                        &   0.973***    &    &   0.705***\\
                &                         &   (0.093)  &     &   (0.098)\\
    $\log(commute\;distance_i)$ &       &           &  0.375***          & 0.327***\\
               	&              &           &(0.041)            &   (0.043)\\
    Const       & -1.338***   &   0.248***    &     -2.526*** & -4.108*** \\
                &  (0.161)     &     (0.052) &    (0.041) &  (0.395) \\ 
    \midrule
    Average marginal effects\\
    \midrule
    $strat_{c(i,work)}$   &     12.5  &           &    &  10.2  \\
    $formal_{c(i,work)}$    &      &   21.3   &    &  14.3\\
    $\log(commute\;distance_i)$ &              &           &  8.3          & 6.6\\
    \midrule
     N       &    2441        &      2441       &     2441         &   2441   \\          
    LL        &      -1507     &    -1533      & -1548       & -1442\\
    \bottomrule
    \addlinespace[1ex]
    \end{tabular}
    \label{table:bogota}
    \end{center}
\end{subtable}

\begin{subtable}{\textwidth}
\begin{center}
\caption{\emph{Low stratum commuters}}
Dep. Variable: $\log\left(\frac{\mathrm{Pr}(disrupt_i=1|i\in Bogot\acute{a})}{1-\mathrm{Pr}(disrupt_i=1\in Bogot\acute{a})}\right)$ \\ 
\begin{tabular}{c c c c c}
\toprule
Ind. variable & \textbf{(I)} & \textbf{(II)}  & \textbf{(III)}& \textbf{(IV)}\\ 

\midrule
$strat_{c(i,work)}$  &     0.541***   &           &     &  0.293***  \\
              & (0.095)               &           &     &   (0.106)\\
$formal_{c(i,work)}$      &                        &  1.196***    &    &   0.779***\\
            &                         &   (0.203)  &     &   (0.220)\\
$\log(commute\;distance_i)$ &       &           &  0.638***          & 0.351***\\
           	&              &           &(0.098)            &   (0.115)\\
Const       & -1.411***   &   -0.097    &     -5.244*** & -3.914*** \\
            &  (0.303)     &     (0.107) &    (0.851) &  (0.887) \\ 
\midrule
Average marginal effects\\
\midrule
$strat_{c(i,work)}$   &     12.4  &           &    &  6.2  \\
$formal_{c(i,work)}$    &      &   27.3   &    &  16.7\\
$\log(commute\;distance_i)$ &              &           &  14.3         & 7.5\\
\midrule
 N       &   531        &     531     &   531        &   531   \\          
LL        &      -344     &    -344      & -340       & -328\\

\bottomrule
\addlinespace[1ex]
\end{tabular}
\label{table:bogota_low}
\end{center}
\end{subtable}

\end{table}

%% file: Discussion.tex
Previous work has shown that lockdown policies are successful at reducing mobility and disease caseloads, using data from Europe \cite{flaxman2020estimating} and China \cite{fang2020human}. Here we consider whether the restrictiveness of the policy in terms of the share of people allowed out per day is correlated with the size of the reduction in mobility. We found, in accordance with the previous literature, that all cities experienced reductions in urban mobility - but there is no statistical relationship between the severity of local mobility quotas and the degree of mobility reduction. Larger, wealthier cities reduced mobility the most, even though they generally imposed less severe mobility quotas. Smaller cities, with more informal employment, did not experience a comparable reduction. 

While the signal linking city size and variables capturing local economic structure to mobility reduction is strong, there are a number of other factors which likely played a role. These include the degree of enforcement of the policy, the availability of economic aid to workers and firms, population density, and the number of infections in the city. For example, on a national level, the government provided cash transfers to informal workers and families before and during the study period \cite{billion}, in addition to a host of economic relief measures \cite{cgd,cgd2}. Future work might investigate further the role and distribution of aid in policies aimed at mobility restriction. Additionally, residents of cities with larger caseloads early in the pandemic are likely to have been more willing to lockdown \cite{van2020using}. Although further investigation is needed, we expect to find that (as with GDP per capita, formality rate and industrial complexity) these factors correlate with city size and are thus consistent with our findings. 

Our findings highlight the role of labor structure in cities' ability to reduce mobility. Less wealthy, often informal firms have less financial capacity to close operations. Workers at these firms are thus incentivized to continue working despite mobility restrictions. The example of Bogot\'a depicts this clearly - residents commuting to high income/formalized work areas were as much as twice as likely to disrupt their commutes (as compared to commutes to low income areas). This result is borne out even for workers from low income residential areas, suggesting that firm closures in these higher income/formalized areas affect both low and high income workers alike. 

Our findings underscore the need for future policy measures to increase aid-based measures alongside or even in place of mobility restrictions. A large share of workers particularly in smaller cities - often informally employed and rarely in teleworkable occupations - cannot work from home (potentially leading to more rapid case growth in smaller cities \cite{ribeiro2020city}), and require substantial and sustained support, potentially leading to faster spread. While many countries encompassing middle and high income economies alike have developed historically unprecedented wage support schemes, the ability of low and middle income countries to sustain such schemes in the longterm is a topic of central interest. This question’s importance is underscored by the current situation in which lower/middle income countries will see their vaccine stockpiles grow more slowly. As countries wrestle with how to balance reopening the economy with persistent reduction of viral spread, policymakers must consider the capacity of the non-teleworking, often informal population to adhere to lockdown measures.

%% file: scifile.bbl
\begin{thebibliography}{10}

\bibitem{hsiang2020effect}
Hsiang S, Allen D, Annan-Phan S, Bell K, Bolliger I, Chong T, et~al.
\newblock The effect of large-scale anti-contagion policies on the COVID-19
  pandemic.
\newblock Nature. 2020;584(7820):262--267.

\bibitem{fang2020human}
Fang H, Wang L, Yang Y.
\newblock Human mobility restrictions and the spread of the novel coronavirus
  (2019-ncov) in china.
\newblock Journal of Public Economics. 2020;191:104272.

\bibitem{maier2020effective}
Maier BF, Brockmann D.
\newblock Effective containment explains subexponential growth in recent
  confirmed COVID-19 cases in China.
\newblock Science. 2020;368(6492):742--746.

\bibitem{flaxman2020estimating}
Flaxman S, Mishra S, Gandy A, Unwin HJT, Mellan TA, Coupland H, et~al.
\newblock Estimating the effects of non-pharmaceutical interventions on
  COVID-19 in Europe.
\newblock Nature. 2020;584(7820):257--261.

\bibitem{sostero2020teleworkability}
Sostero M, Milasi S, Hurley J, Fernandez-Macias E, Bisello M.
\newblock Teleworkability and the COVID-19 crisis: a new digital divide?
\newblock European Commission: Joint Research Cente (JRC) Working Papers Series
  on Labour, Education and Technology; 2020.

\bibitem{brussevich2020will}
Brussevich M, Dabla-Norris E, Khalid S. Who will Bear the Brunt of Lockdown
  Policies? Evidence from Tele-workability Measures Across Countries.
\newblock IMF Working Paper; 2020.

\bibitem{bonaccorsi2020economic}
Bonaccorsi G, Pierri F, Cinelli M, Flori A, Galeazzi A, Porcelli F, et~al.
\newblock Economic and social consequences of human mobility restrictions under
  COVID-19.
\newblock Proceedings of the National Academy of Sciences.
  2020;117(27):15530--15535.

\bibitem{fraiberger2020uncovering}
Fraiberger SP, Astudillo P, Candeago L, Chunet A, Jones NK, Khan MF, et~al.
\newblock Uncovering socioeconomic gaps in mobility reduction during the
  COVID-19 pandemic using location data.
\newblock arXiv preprint arXiv:200615195. 2020;.

\bibitem{carvalho2020tracking}
Carvalho VM, Hansen S, Ortiz A, Garcia JR, Rodrigo T, Rodriguez~Mora S, et~al.
\newblock Tracking the Covid-19 crisis with high-resolution transaction data.
\newblock BBVA Research Working Paper. 2020;.

\bibitem{LUXURY}
Valentino-DeVries J, Dance GJ.
\newblock Location data says it all: Staying at home during coronavirus is a
  luxury.
\newblock New York Times. 2020;Available from:
  \url{https://www.nytimes.com/interactive/2020/04/03/us/coronavirus-stay-home-rich-poor.html}.

\bibitem{ONS_data}
Caul S.
\newblock Deaths involving COVID-19 by local area and socioeconomic
  deprivation: deaths occurring between 1 March and 31 July 2020.
\newblock UK Office for National Statistics; 2020.

\bibitem{zhang2020spatial}
Zhang CH, Schwartz GG.
\newblock Spatial disparities in coronavirus incidence and mortality in the
  United States: an ecological analysis as of May 2020.
\newblock The Journal of Rural Health. 2020;36(3):433--445.

\bibitem{lancet2020covid}
Prado B.
\newblock COVID-19 in Brazil:“So what?”.
\newblock Lancet. 2020;395(1461):31095--3.

\bibitem{pc_spanish}
Guerrero M.
\newblock 'Pico y cédula' en Colombia: ciudades donde aplica la medida.
\newblock El Tiempo. 2020;Available from:
  \url{https://www.eltiempo.com/economia/empresas/pico-y-cedula-en-colombia-conozca-en-que-ciudades-aplica-la-medida-480328}.

\bibitem{pc_english}
Alsema A.
\newblock Coronavirus information for foreigners in Colombia.
\newblock Colombia Reports. 2020;Available from:
  \url{https://colombiareports.com/coronavirus-information-for-immigrants/}.

\bibitem{duranton2016agglomeration}
Duranton G.
\newblock Agglomeration effects in Colombia.
\newblock Journal of Regional Science. 2016;56(2):210--238.

\bibitem{lora2016path}
O'Clery N, Gomez-Lievano A, Lora E.
\newblock The path to labor formality: Urban agglomeration and the emergence of
  complex industries.
\newblock Center for International Development at Harvard University; 2016.

\bibitem{peak2018population}
Peak CM, Wesolowski A, zu~Erbach-Schoenberg E, Tatem AJ, Wetter E, Lu X, et~al.
\newblock Population mobility reductions associated with travel restrictions
  during the Ebola epidemic in Sierra Leone: use of mobile phone data.
\newblock International journal of epidemiology. 2018;47(5):1562--1570.

\bibitem{badr2020association}
Badr HS, Du H, Marshall M, Dong E, Squire MM, Gardner LM.
\newblock Association between mobility patterns and COVID-19 transmission in
  the USA: a mathematical modelling study.
\newblock The Lancet Infectious Diseases. 2020;.

\bibitem{xiong2020mobile}
Xiong C, Hu S, Yang M, Luo W, Zhang L.
\newblock Mobile device data reveal the dynamics in a positive relationship
  between human mobility and COVID-19 infections.
\newblock Proceedings of the National Academy of Sciences.
  2020;117(44):27087--27089.

\bibitem{Gozzi2020.10.08.20204750}
Gozzi N, Tizzoni M, Chinazzi M, Ferres L, Vespignani A, Perra N.
\newblock Estimating the effect of social inequalities in the mitigation of
  COVID-19 across communities in Santiago de Chile.
\newblock medRxiv. 2020;Available from:
  \url{https://www.medrxiv.org/content/early/2020/10/13/2020.10.08.20204750}.

\bibitem{glaeser2020much}
Glaeser EL, Gorback CS, Redding SJ.
\newblock How much does covid-19 increase with mobility? evidence from New York
  and four other US cities.
\newblock National Bureau of Economic Research; 2020.

\bibitem{gauvin2020gender}
Gauvin L, Tizzoni M, Piaggesi S, Young A, Adler N, Verhulst S, et~al.
\newblock Gender gaps in urban mobility.
\newblock Humanities and Social Sciences Communications. 2020;7(1):1--13.

\bibitem{woskie2020men}
Woskie LR, Wenham C.
\newblock Do men and women lockdown differently? An examination of panamas
  COVID-19 sex-segregated social distancing policy.
\newblock medRxiv. 2020;.

\bibitem{louail2015uncovering}
Louail T, Lenormand M, Picornell M, Cant{\'u} OG, Herranz R, Frias-Martinez E,
  et~al.
\newblock Uncovering the spatial structure of mobility networks.
\newblock Nature communications. 2015;6(1):1--8.

\bibitem{lotero2016rich}
Lotero L, Hurtado RG, Flor{\'\i}a LM, G{\'o}mez-Garde{\~n}es J.
\newblock Rich do not rise early: spatio-temporal patterns in the mobility
  networks of different socio-economic classes.
\newblock Royal Society Open Science. 2016;3(10):150654.

\bibitem{alexander2015origin}
Alexander L, Jiang S, Murga M, Gonz{\'a}lez MC.
\newblock Origin--destination trips by purpose and time of day inferred from
  mobile phone data.
\newblock Transportation Research Part C: Emerging technologies.
  2015;58:240--250.

\bibitem{wesolowski2013impact}
Wesolowski A, Eagle N, Noor AM, Snow RW, Buckee CO.
\newblock The impact of biases in mobile phone ownership on estimates of human
  mobility.
\newblock Journal of the Royal Society Interface. 2013;10(81):20120986.

\bibitem{gsma}
GSMA. The mobile economy: Latin America; 2020.
\newblock
  https://www.gsma.com/mobileeconomy/wp-content/uploads/2020/12/GSMA{\_}MobileEconomy2020{\_}LATAM{\_}Eng.pdf.

\bibitem{pappalardo2020individual}
Pappalardo L, Ferres L, Sacasa M, Cattuto C, Bravo L.
\newblock An individual-level ground truth dataset for home location detection.
\newblock arXiv preprint arXiv:201008814. 2020;.

\bibitem{flowminder}
Flowminder. Understanding CDR aggregates: Fundamentals; 2020.
\newblock
  https://covid19.flowminder.org/cdr-aggregates/understanding-cdr-aggregates-fundamentals.

\bibitem{scikit-learn}
Pedregosa F, Varoquaux G, Gramfort A, Michel V, Thirion B, Grisel O, et~al.
\newblock Scikit-learn: Machine Learning in {P}ython.
\newblock Journal of Machine Learning Research. 2011;12:2825--2830.

\bibitem{ahas2010using}
Ahas R, Silm S, J{\"a}rv O, Saluveer E, Tiru M.
\newblock Using mobile positioning data to model locations meaningful to users
  of mobile phones.
\newblock Journal of Urban Technology. 2010;17(1):3--27.

\bibitem{vanhoof2018assessing}
Vanhoof M, Reis F, Ploetz T, Smoreda Z.
\newblock Assessing the quality of home detection from mobile phone data for
  official statistics.
\newblock Journal of Official Statistics. 2018;34(4):935--960.

\bibitem{escape}
Marsh S.
\newblock Escape to the country: how Covid is driving an exodus from
  Britain’s cities.
\newblock The Guardian. 2020;Available from:
  \url{https://www.theguardian.com/world/2020/sep/26/escape-country-covid-exodus-britain-cities-pandemic-urban-green-space}.

\bibitem{DANE}
{Departamento Administrativo Nacional de Estadística (DANE)}. Censo nacional
  de población y vivienda; 2020.
\newblock
  https://www.dane.gov.co/index.php/en/estadisticas-por-tema/demografia-y-poblacion/censo-nacional-de-poblacion-y-vivenda-2018.

\bibitem{2020SciPy-NMeth}
Virtanen P, Gommers R, Oliphant TE, Haberland M, Reddy T, Cournapeau D, et~al.
\newblock {{SciPy} 1.0: Fundamental Algorithms for Scientific Computing in
  Python}.
\newblock Nature Methods. 2020;17:261--272.

\bibitem{datlas}
{Colombian Atlas of Economic Complexity}. Colombian formal employment and
  industry/region mapping.
\newblock The Center for International Development (CID) at Harvard University;
  2017.
\newblock http://datlascolombia.com/.

\bibitem{hidalgo2009building}
Hidalgo CA, Hausmann R.
\newblock The building blocks of economic complexity.
\newblock Proceedings of the National Academy of Sciences.
  2009;106(26):10570--10575.

\bibitem{hidalgo2007product}
Hidalgo CA, Klinger B, Barab{\'a}si AL, Hausmann R.
\newblock The product space conditions the development of nations.
\newblock Science. 2007;317(5837):482--487.

\bibitem{kraemer2020effect}
Kraemer MU, Yang CH, Gutierrez B, Wu CH, Klein B, Pigott DM, et~al.
\newblock The effect of human mobility and control measures on the COVID-19
  epidemic in China.
\newblock Science. 2020;368(6490):493--497.

\bibitem{bettencourt2007growth}
Bettencourt LM, Lobo J, Helbing D, K{\"u}hnert C, West GB.
\newblock Growth, innovation, scaling, and the pace of life in cities.
\newblock Proceedings of the National Academy of Sciences.
  2007;104(17):7301--7306.

\bibitem{bettencourt2010urban}
Bettencourt LM, Lobo J, Strumsky D, West GB.
\newblock Urban scaling and its deviations: Revealing the structure of wealth,
  innovation and crime across cities.
\newblock PloS one. 2010;5(11):e13541.

\bibitem{davis2020comparative}
Davis DR, Dingel JI.
\newblock The comparative advantage of cities.
\newblock Journal of International Economics. 2020;123:103291.

\bibitem{brinkman2014supply}
Brinkman J. The supply and demand of skilled workers in cities and the role of
  industry composition.
\newblock FRB of Philadelphia Working Paper; 2014.

\bibitem{alves2015scale}
Alves LG, Mendes RS, Lenzi EK, Ribeiro HV.
\newblock Scale-adjusted metrics for predicting the evolution of urban
  indicators and quantifying the performance of cities.
\newblock PloS one. 2015;10(9):e0134862.

\bibitem{weill2020social}
Weill JA, Stigler M, Deschenes O, Springborn MR.
\newblock Social distancing responses to COVID-19 emergency declarations
  strongly differentiated by income.
\newblock Proceedings of the National Academy of Sciences.
  2020;117(33):19658--19660.

\bibitem{ribeiro2020city}
Ribeiro HV, Sunahara AS, Sutton J, Perc M, Hanley QS.
\newblock City size and the spreading of COVID-19 in Brazil.
\newblock PloS one. 2020;15(9):e0239699.

\bibitem{roberts2005globalization}
Roberts BR.
\newblock Globalization and Latin American cities.
\newblock International Journal of Urban and Regional Research.
  2005;29(1):110--123.

\bibitem{medina2007stratification}
Medina C, Morales L, Bernal R, Torero M.
\newblock Stratification and Public Utility Services in Colombia: Subsidies to
  Households or Distortion of Housing Prices?
\newblock Economia. 2007;7(2):41--99.

\bibitem{dingel2020many}
Dingel JI, Neiman B.
\newblock How many jobs can be done at home?
\newblock National Bureau of Economic Research; 2020.

\bibitem{hatayama2020jobs}
Hatayama M, Viollaz M, Winkler H.
\newblock Jobs' Amenability to Working from Home: Evidence from Skills Surveys
  for 53 Countries.
\newblock World Bank Policy Research Working Paper. 2020;(9241).

\bibitem{chetty2020did}
Chetty R, Friedman JN, Hendren N, Stepner M, et~al.
\newblock How did COVID-19 and stabilization policies affect spending and
  employment? A new real-time economic tracker based on private sector data.
\newblock National Bureau of Economic Research; 2020.

\bibitem{guzman2017city}
Guzman LA, Oviedo D, Bocarejo JP.
\newblock City profile: The bogot{\'a} metropolitan area that never was.
\newblock Cities. 2017;60:202--215.

\bibitem{billion}
Symmes~Cobb J.
\newblock Colombia to spend 3.65 bln on coronavirus economic measures.
\newblock Thomson Reuters Foundation. 2020;Available from:
  \url{https://uk.reuters.com/article/health-coronavirus-colombia-economy/colombia-to-spend-3-65-bln-on-coronavirus-economic-measures-idUSL1N2BB0VX}.

\bibitem{cgd}
Ravallion M.
\newblock Pandemic Policies in Poor Places.
\newblock Center for Global Development; 2020.

\bibitem{cgd2}
Cárdenas M, Martínez~Beltrán H.
\newblock COVID-19 in Colombia: Impact and Policy Responses.
\newblock Center for Global Development; 2020.

\bibitem{van2020using}
Van~Bavel JJ, Baicker K, Boggio PS, Capraro V, Cichocka A, Cikara M, et~al.
\newblock Using social and behavioural science to support COVID-19 pandemic
  response.
\newblock Nature Human Behaviour. 2020;p. 1--12.

\end{thebibliography}
